\documentclass[11pt]{article}

\usepackage{apacite}
\usepackage{amsmath}
\usepackage{amsfonts}
\usepackage{color}

\usepackage{txfonts}
\usepackage[scaled=.9]{helvet}

\usepackage{epsfig}
\usepackage{epstopdf} 
\usepackage[tight]{subfigure}
\usepackage{url}
\usepackage{xspace}
\usepackage{multirow}
\usepackage{booktabs}
\usepackage[left=1.8cm,right=1.8cm, top=2cm, bottom=2cm]{geometry}

\newcommand{\paratitle}[1]{\vspace{1.5ex}\noindent \textbf{#1}}
\newcommand{\ie}{\emph{i.e.,}\xspace}
\newcommand{\eg}{\emph{e.g.,}\xspace}
\newcommand{\etal}{\emph{et al.}\xspace}

\newcommand{\eat}[1]{}

\begin{document}

\title{NEXT: A Neural Network Framework for Next POI Recommendation}

\author{
Zhiqian Zhang$^a$, Chenliang Li$^a$, Zhiyong Wu$^a$, Aixin Sun$^b$, Dengpan Ye$^a$, Xiangyang Luo$^c$\\ $^a$State Key Lab of Software Engineering, Wuhan University, China 430079\\
\texttt{E-mail: \{cllee,zhangzq2011,zywu\}@whu.edu.cn, yedp2001@163.com}
\\ $^b$School of Computer Engineering, Nanyang Technological University, Singapore 639798\\
\texttt{E-mail: axsun@ntu.edu.sg}
\\ $^c$State Key Lab of Mathematical Engineering and Advanced Computing, China 450001\\
\texttt{E-mail: xiangyangluo@126.com}
\\
}

\date{}

\maketitle

\begin{abstract}
The task of \textit{next POI recommendation} has been studied extensively in recent years. However, developing an unified recommendation framework to incorporate multiple factors associated with both POIs and users remains challenging, because of the heterogeneity nature of these information. Further, effective mechanisms to handle cold-start and endow the system with interpretability are also difficult topics. Inspired by the recent success of neural networks in many areas, in this paper, we present a simple but effective neural network framework for next POI recommendation, named NEXT. NEXT is an unified framework to learn the hidden intent regarding user's next move, by incorporating different factors in an unified manner. Specifically, in NEXT, we incorporate  meta-data information and two kinds of temporal contexts (\ie time interval and visit time). To leverage sequential relations and geographical influence, we propose to adopt DeepWalk, a network representation learning technique, to encode such knowledge. We evaluate the effectiveness of NEXT against state-of-the-art alternatives and neural networks based solutions. Experimental results over three publicly available datasets demonstrate that NEXT significantly outperforms baselines in real-time next POI recommendation. Further experiments demonstrate the superiority of NEXT in handling cold-start. More importantly, we show that NEXT  provides meaningful explanation of the dimensions in hidden intent space.
\end{abstract}

\section{Introduction}\label{sec:intro}

The huge volume of check-in data from various location-based social networks (LBSNs) enables studies on human mobility behavior in a large scale. Next POI recommendation, the task to predict the next POI a user will visit at a specific time point given her historical check-in data,  has been studied extensively in recent years.

Location recommendation is different from a typical recommendation task (\eg movies, songs, books) because there are a wide range of contextual factors  to consider. These auxiliary factors include the temporal context, sequential relations, geographical influence, and auxiliary meta-data information (such as textual description, user friendship, to name a few). Existing solutions based on matrix factorization and embedding learning techniques have delivered encouraging performance. However, these approaches often incorporate different contextual factors through a fusion strategy: modeling the factors either as weighting coefficients or additional constraints~\cite{ijcai13:cheng,ijcai15:feng,gis10:ye,sigir11:ye,sigir13:yuan}. The dense vector representation (\ie embeddings) and neural networks (NN) based techniques provide a new way of modeling these factors in an unified manner. This offers two benefits:
\begin{itemize}
  \item In contrast to one-hot representation in high dimensional space, dense representation enables us to retain the semantic relatedness or constraints in the embeddings of POIs, users and their auxiliary meta-data. For example, users at \textit{Golden Gate Overlook} are likely to visit \textit{Baker Beach} in San Francisco, and vice versa. Therefore, \textit{Golden Gate Overlook} and \textit{Baker Beach} can be projected closer in the embedding space. By encoding all context factors mentioned above into corresponding embeddings or nonlinear manipulations, we expect to obtain better recommendation accuracy.
  \item The cold-start problem can be further alleviated, since both new users and POIs may be covered partially by some of the factors (\eg textual description, friendship). With dense vector representations, we can approximate these freshers in a smooth manner.
\end{itemize}
Although the outlook is encouraging, however, the challenge is how to utilize these context factors effectively in NN, and also achieve an explainable model based on the auxiliary meta-data.

In this paper, we take a special interest on developing an unified neural network based framework to address the above challenge. We propose a simple but effective \textbf{n}eural n\textbf{e}twork framework for ne\textbf{xt} POI recommendation task, named NEXT. With one layer of feed-forward neural network supercharged by ReLU (\ie rectified linear unit), NEXT  is able to incorporate temporal context, sequential relations, geographical influence and auxiliary meta-data information, in an integrated architecture.

Specifically, NEXT utilizes one-layer of nonlinearity to learn high-level spatial intent for a user from both the user and her latest POI visit. In other words, NEXT does not  calculate an inner product directly on the embeddings of users and POIs in a common hidden space as in the existing embedding learning approaches~\cite{ijcai15:feng,sigir15:xutao,cikm16:xie}. Instead, NEXT utilizes the non-linear transformations to extract the user-based and POI-based spatial intents respectively. Empowered by this separation and nonlinearity extraction, we can incorporate temporal context, auxiliary meta-data information into the user-based and POI-based intent learning process  in an integrated manner. This design  also enables us to tackle the cold-start problem and deliver semantic interpretation for each individual intent dimension. To further leverage the sequential relations and geographical influence in the context of POI recommendation without feature engineering, we devise a strategy to pre-train POI embeddings. The resultant POI embeddings encode the sequential relations and geographical influence.

Based on three real-world datasets, the proposed NEXT achieves significantly better recommendation performance than existing state-of-the-art approaches and neural networks based alternatives. In summary, the main contributions of this paper are as follows:
\begin{itemize}
\item We present a novel neural network based solution for the task of next POI recommendation. The proposed NEXT is an unified framework such that temporal context, sequential relations, geographical influence and auxiliary meta-data information can be exploited naturally. By injecting meta-data information into the intent learning process, we endow NEXT with the ability to smoothly handle cold-start recommendation.
\item Given the textual meta-data, we show that NEXT enables the interpretable hidden feature learning for  explaining the recommendations. This in turn potentially benefits cold-start.
\item We adopt the network representation learning technique to pre-train POI embeddings. This pre-training strategy enables us to retain the sequential relations and geographical influence for better model learning. This is a flexible strategy such that other constraints besides these two context factors can also be captured.
\end{itemize}

\section{Related Work}\label{sec:related}
Our work is related to two lines of literatures, POI recommendation and neural networks. We review the recent advances in both areas.

\subsection{POI Recommendation}
The conventional collaborative filtering (CF) techniques have been widely studied for POI recommendation~\cite{gis10:ye,sigir11:ye,sigir13:yuan}. Ye~\etal proposed a friend-based collaborative filtering (FCF) approach for POI recommendation based on common visited POIs of the friends~\cite{gis10:ye}. Temporal context information and geographical constraints were then proven to be effective for  recommendation~\cite{sigir11:ye,sigir13:yuan,tois16:yin}.

Recently, recommendation models based on matrix factorization and embedding learning have been intensively studied. Cheng~\etal proposed a multi-center Gaussian model to capture user geographical influence and combined it with matrix factorization model to recommend POIs~\cite{aaai12:cheng}. In~\cite{ijcai13:cheng},  a tensor-based FPMC-LR model is proposed by considering the first-order Markov chain for POI transitions and distance constraints.  Li~\etal proposed a ranking based factorization method for POI recommendation which learns factorization by fitting the user's preference over POIs, where the preference was measured in terms of POI visit frequency~\cite{sigir15:xutao}. Feng~\etal integrated sequential information, individual preference and geographical influence into a personalized ranking metric embedding model to improve the recommendation performance~\cite{ijcai15:feng}. Gao~\etal introduced matrix factorization based POI recommendation algorithm with temporal influence based on two temporal properties: non-uniforms and consecutiveness~\cite{cikm12:gao}. He~\etal proposed a tensor-based latent model which incorporates the date information, geographical distance and personal POI transition patterns into an unified framework~\cite{aaai16:he}. Zhao~\etal developed a ranking-based pairwise tensor factorization framework, named STELLAR~\cite{aaai16:zhao}. STELLAR incorporates fine-grained temporal contexts (\ie month, weekday/weekend and hour) and brings the significant improvement. These works tried to fit the model by maximizing the interaction between users and POIs, where the recommendation decision is made based on the last POI visit alone. Recently, Xie~\etal proposed an embedding learning approach that utilizes a bipartite graph to model a pair of context factors in the context of POI recommendation, named GE model~\cite{cikm16:xie}. Four pairs of context factors: POI-POI, POI-Region, POI-Time, POI-Word were modeled in an unified optimization framework. The experimental results showed that GE significantly outperforms other alternative algorithms for real-time next POI recommendation.

\subsection{Neural Networks}
Neural networks techniques have experienced great success in  natural language processing area such as language modeling~\cite{interspeech10:mikolov,arxiv13:mikolov}, machine translation~\cite{corr14:cho,corr14:bahdanau}, question answering~\cite{acl16:wang}, summarization~\cite{icml16:allamanis}, etc. The conventional neural networks such as artificial neural network (ANN)~\cite{ditc1986:rumelhart} and multilayer perceptron (MLP) architectures~\cite{ditc1986:rumelhart,werbos1988,bishop1995} are  among the first invented networks. Although relatively simple, it has been proven that a MLP with a single hidden layer containing a sufficient number of nonlinear units can approximate any continuous function on a compact input domain to arbitrary precision~\cite{hornik1989}. Recently, He~\etal developed a deep neural network based matrix factorization approach for collaborative filtering with implicit feedback data~\cite{www17:he}. Based on the embeddings of items and users, they applied multiple layers of MLP to extract the high-level hidden feature by maximizing user-item interactions.

Among the various neural network structures, recurrent neural networks (RNN) have been widely used to model sequential data of arbitrary length with its recurrent calculation of hidden representation~\cite{cs90:elman,interspeech10:mikolov}. RNN has been successfully adopted in the tasks like poem generation~\cite{ijcai16:yan} and sequential click prediction~\cite{aaai14:zhang}. However, RNN suffers from the \textit{exploding or vanishing gradients} problem~\cite{nc97:sepp} such that the distant dependencies within the longer sequence could not be learnt appropriately. Two RNN variants: long short-term memory (LSTM) and gated recurrent unit (GRU), were proposed to tackle this problem to enable long-term dependency learning. LSTM utilizes three gates and a memory cell to control the information flow. It forgets the irrelevant signals by turning off the corresponding three gates and updating memory content. LSTM has been widely used in different tasks involving the sequence modeling~\cite{emnlp15:chen,corr15:tim}. GRU is a recent variant of RNN with two gates and no memory cell~\cite{corr14:cho}. The two gates control the expose of the previous hidden output and the update of the new hidden output respectively. GRU has been proven to capture the long-term dependencies just like LSTM~\cite{corr14:chung}. There is very limited studies on using neural network for next POI recommendation. Liu~\etal proposed a RNN-based neural network solution by modeling user's historical POI visits in a sequential manner~\cite{aaai16:liu}, named STRNN. STRNN adopts time-specific transition matrices and distance-specific transition matrices in a recurrent manner under the framework of RNN model. The proposed NEXT here differs significantly from STRNN in several aspects. First, NEXT is one layer feed-forward neural network based model where only the latest POI visit is taken as input. On the contrary, STRNN (and also other RNN variants) has to take all historical POI visits as input and process in a sequential (or recurrent) manner, which increases the complexity of the model. Second, while STRNN only incorporates temporal context and geographical influence for recommendation, NEXT can incorporate multiple context factors (\ie temporal context, geographical influence, auxiliary meta-data) in an unified framework. Third, instead of applying distance-specific latent feature extraction in STRNN, NEXT encodes the sequential relations (transition behaviors and geographical information) within the pre-trained POI embeddings  by adopting DeepWalk technique~\cite{kdd14:perozzi}. Our experimental results show that NEXT delivers superior performance than the existing matrix factorization and embedding learning based models as well as the neural network based techniques.

\section{Our Approach}\label{sec:algo}
In this section, we first formally define the research problem and then present the proposed \textbf{n}eural n\textbf{e}twork framework for ne\textbf{xt} POI recommendation task, named NEXT. We introduce the basic neural architecture of NEXT to extract the hidden intents regarding the user's next move. We then describe the mechanism to accommodate NEXT with the temporal context modeling. Next, a pre-training strategy based on the network representation technique (\ie DeepWalk) is introduced to integrate the sequential relations and geographical influence. We also discuss the traits of NEXT to interpret the hidden intent features and handle the cold-start issue.

\subsection{Next POI Recommendation}\label{ssec:nextpoi}

We first  define the problem of next POI recommendation.  Given a user with a sequence of historical POI visits $L^u_i=\{q^u_{t_1},q^u_{t_2},...,q^u_{t_{i-1}}\}$ up to time $t_{i-1}$, the task is to calculate a score for each POI based on $L^u_i$ and a time point $t_i$. Higher score indicates higher probability that the user will like to visit that POI at time $t_i$. The POI with the highest score will then be recommended.

In most LBSNs, in addition to the sequence of historical POI visits, users and POIs are associated with auxiliary meta-data information. For example, a user could build connections with other users (\eg friends) to share their activities and opinions. A POI could contain textual  description or category labels. Here, we denote the auxiliary meta-data associated with user $u$ and POI $q$ as $\mathcal{A}_u$ and $\mathcal{A}_q$, respectively.

\subsection{Neural Architecture}\label{ssec:architecture}

\paratitle{Basic Model.}  Different from the existing works that directly take the shallow embeddings of the users and POIs for score calculation (\ie an inner product), in NEXT, we introduce an additional feed-forward neural network layer to model user's spatial intent, on top of the embedding.

Let $\mathbf{u}_u\in R^d$ be the embedding of user $u$, $\mathbf{q}_\ell\in R^d$ be the embedding of a candidate POI $q_\ell$ to be recommended, and $\mathbf{q}^u_{t_{i-1}}\in R^d$ be the embedding of POI $q^u_{t_{i-1}}$, the last visited POI by user $u$ at time $t_{i-1}$.\footnote{For model simplicity, we set intent vectors, POI embeddings, user embeddings, word embeddings to be of the same dimension.} We model the hidden intent of next visit by a nonlinear activation function, \textit{rectified linear unit}: $ReLU(x)=\max(x,0)$.
\begin{align}
\mathbf{h}^q_{t_{i}}&=ReLU(\mathbf{W}_1\mathbf{q}^u_{t_{i-1}}+\mathbf{b}_1)\label{eqn:poi}\\
\mathbf{h}^u&=ReLU(\mathbf{W}_2\mathbf{u}_u+\mathbf{b}_2)\label{eqn:user}\\
\mathbf{c}_{\ell}&=ReLU(\mathbf{W}_3\mathbf{q}_\ell+\mathbf{b}_3)\label{eqn:candidate}
\end{align}
In the above modeling, the hidden intent vector $\mathbf{h}^q_{t_{i}}$ is expected to capture semantics regarding the user's next move at time $t_{i}$ based on her last POI visit. Intent vector  $\mathbf{h}^u$ captures  user specific knowledge on spatial preference of a particular user. $\mathbf{c}_\ell$ is the intent representation of candidate POI $\ell$.  $\mathbf{W}_1\in R^{d\times d}$  and $\mathbf{W}_2\in R^{d\times d}$ are  transition matrices from  POI embeddings and user embeddings respectively, to the hidden intent.  $\mathbf{W}_3\in R^{d\times d}$ is a weight matrix. $\mathbf{b}_1\in R^d$, $\mathbf{b}_2\in R^d$, and $\mathbf{b}_3\in R^d$ are all bias vectors.

With the hidden intent vectors $\mathbf{h}^q_{t_{i}}$, $\mathbf{h}^u$, and $\mathbf{c}_{\ell}$, the recommendation score $y_{u,t_i,\ell}$ of POI $q_\ell$ for user $u$ at time $t_{i}$ is computed as follows:
\begin{equation}
y_{u,t_i,\ell}=(\mathbf{h}^{u}+\mathbf{h}_{t_i}^q)^T\mathbf{c}_{\ell}\label{eqn:score}
\end{equation}

\begin{figure}{\includegraphics[width=.65\linewidth] {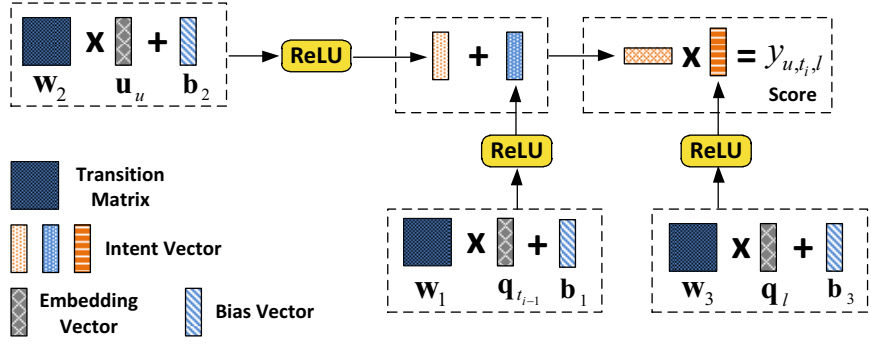}
 }%
 \centering
\caption{Basic model of NEXT, $\mathbf{u}_u$, $\mathbf{q}_\ell$ and $\mathbf{q}_{t_{i-1}}$ are the embedding vectors of the user, candidate poi, and the last visited POI.}
\label{fig:basic}
\end{figure}

In simple words, in NEXT, instead of directly using embedding vectors of users and POIs,  a feed-forward network layer is used to transform the embeddings to intent vectors. Recommendations are made based on the intent vectors. The transition matrices and bias vectors make it possible to identify the most useful information from the embeddings. By separating the intent vectors and embedding vectors,  NEXT framework also makes it simple and straightforward to be extended by incorporating information from different context factors. Figure~\ref{fig:basic} illustrates the basic model of NEXT.

\eat{
Equation~\ref{eqn:candidate} is a feed-forward network layer that transforms the POI embedding of candidate POI $\ell$ into the hidden intent space.

Specifically, we choose to separately infer the spatial intents from a user and her last visited POI, respectively. Given the last POI visit $q^u_{t_{i-1}}$ happened at time $t_{i-1}$, the hidden intent vector $\mathbf{h}^q_{t_{i}}$ of user $u$ at time $t_{i}$ can be calculated by NEXT as follows:

where $\mathbf{q}^u_{t_{i-1}}\in R^d$ is the embedding of POI $q^u_{t_{i-1}}$,

In the above formulation, the hidden intent $\mathbf{h}^q_{t_{i}}$ is derived solely based on a single POI alone. Therefore it is a static information for all users under consideration. On the other hand, user specific knowledge is expected to contain spatial preference of a particular user. Similar to Equation~\ref{eqn:poi},  we derive user-specific spatial intent as follow:
\begin{equation}
\mathbf{h}^u=ReLU(\mathbf{W}_2\mathbf{u}_u+\mathbf{b}_2)\label{eqn:user}
\end{equation}
where $\mathbf{u}_u\in R^d$ is the embedding of user $u$,
}

\paratitle{Incorporating Meta-data Information.} Since the associated meta-data information could offer complementary knowledge about  users and POIs respectively, it is expected to enhance the understanding of user movement by taking $\mathcal{A}_u$ and $\mathcal{A}_q$ into consideration. Hence, we further enrich NEXT framework by taking these auxiliary semantics into the intent calculations. First, we calculate the embedding $\mathbf{m}_q$ to represent auxiliary meta-data $\mathcal{A}_q$ as follows:
\begin{align}
\mathbf{m}_q=\frac{1}{|\mathcal{A}_q|}\sum_{m\in \mathcal{A}_q}\mathbf{m}_m\label{eqn:metapoi}
\end{align}
where $\mathbf{m}_m$ is the embedding of item $m$ in the meta-data $\mathcal{A}_q$. Based on $\mathbf{m}_q$ from Equation~\ref{eqn:metapoi}, we rewrite Equations~\ref{eqn:poi} and~\ref{eqn:candidate} as follows:
\begin{align}
\mathbf{h}^q_{t_{i}}&=ReLU\left(\mathbf{W}_1\left(\alpha\mathbf{q}^u_{t_{i-1}}+(1-\alpha)\mathbf{m}_{q^u_{t_{i-1}}}\right)+\mathbf{b}_1\right)\label{eqn:intentpoimeta}\\
\mathbf{c}_\ell&=ReLU\left(\mathbf{W}_3\left(\alpha\mathbf{q}_\ell+(1-\alpha)\mathbf{m}_{q_\ell}\right)+\mathbf{b}_3\right)\label{eqn:candidatemeta}
\end{align}
where $\alpha$ works as a tuning parameter, controlling the importance of meta-data information. Similar to Equations~\ref{eqn:metapoi} and~\ref{eqn:intentpoimeta}, we rewrite Equation~\ref{eqn:user} with the auxiliary meta-data $\mathcal{A}_u$ as follows:
\begin{align}
\mathbf{m}_u&=\frac{1}{|\mathcal{A}_u|}\sum_{m\in \mathcal{A}_u}\mathbf{m}_m\label{eqn:metauser}\\
\mathbf{h}^u&=ReLU\big(\mathbf{W}_2\left(\beta\mathbf{u}_u+(1-\beta)\mathbf{m}_u\right)+\mathbf{b}_2\big)\label{eqn:intentusermeta}
\end{align}
where $\mathbf{m}_m$ is the embedding of item $m$ in the meta-data $\mathcal{A}_u$, $\beta$ is a tuning parameter just like $\alpha$ in Equation~\ref{eqn:intentpoimeta}.

Note that the embeddings of users (\ie $\mathbf{u}_u$) and the embeddings of POIs (\ie $\mathbf{q}_q$) are not assumed to be within the same hidden space.  In this sense, given the types of meta-data information are homogenous for $\mathcal{A}_u$ and $\mathcal{A}_q$, NEXT is flexible  to associate two sets of embeddings for the meta-data information. This is reasonable because these two kinds of meta-data may convey very different semantics. For example, both users and POIs can be associated with textual labels. While the users use labels to indicate their tastes and preferred locations, the labels of POIs may cover the related services instead. In this case, we may prefer using two separate embedding spaces.\footnote{We leave the exploration as a part of our future work.}


\subsection{Incorporating Temporal Context}\label{ssec:temporal}
Temporal context has been widely used in existing POI recommendation studies and proven to be effective. Here, we accommodate NEXT with temporal context by influencing the computation of the hidden intent.

There are two kinds of temporal context available: (i) the time interval between two successive POI visits (\ie $t_i-t_{i-1}$), and (ii) the particular time point of the POI visit (\ie $t_i$). For example, a POI visit happened 12 hours ago could contain less guidance about the user's current spatial intent. Similarly, users could express different spatial intents at different time slots, \eg lunch hours. We design a mechanism to incorporate both kinds of temporal context into the POI based intent calculation (Equation~\ref{eqn:intentpoimeta}).

The time interval from the last POI visit is critical to decide the user's next move. However, it is inappropriate to discretize temporal dimension since it is a continuous metric. It is intuitive that the historical POI visits with different time intervals could contain varying spatial intents. And the interplay between the intent and time interval could be complicated and subtle. Here we replace $\mathbf{W}_1$ in Equation~\ref{eqn:poi} with a time interval $t$ dependent transition matrix $\mathbf{W}_{\pi}(t)$ as follows:
\begin{align}
  \mathbf{W}_\pi(t) &= \left\{
\begin{aligned}
&\frac{\pi-t}{\pi}\mathbf{W}_0+\frac{t}{\pi}\mathbf{W}_{\pi},&~\textrm{for~}t<\pi\\
&\mathbf{W}_{\pi},&~\textrm{for~}t>\pi\\
\end{aligned}
\right.\label{eqn:Wpi}
\end{align}
where $\mathbf{W}_0,\mathbf{W}_{\pi}\in R^{n\times d}$ are two transition matrices, $\pi$ is an interval threshold. Equation~\ref{eqn:Wpi} adopts a linear interpolation between $\mathbf{W}_0$ and $\mathbf{W}_{\pi}$ to derive the interval dependent transition matrix. When time interval $t$ is close to $0$, $\mathbf{W}_0$ is mainly in charge of intent calculation, otherwise, $\mathbf{W}_{\pi}$  leads the computation when $t$ approaches $\pi$. $\pi$ works as a window,  and $\mathbf{W}_{\pi}$ is used only when the time interval is larger than $\pi$.

As to the visit time information, we split a day into $24$ time slots, each of which spans one hour (\eg 17:00 - 18:00). Each time slot is associated with a specific bias vector $\mathbf{b_{t}}$. Assigning each time slot with a specific bias vector is reasonable, because  users generally express different POI preferences in different time slots~\cite{sigir13:yuan}. For example, users at the time slots of 20:00 - 22:00 prefer public entertainment. The bias vector for each time slot is expected to store such preference information and correct the mistake incurred by considering the last visited POI alone. For example, an user goes from office to a restaurant. If this transition happens in the midnoon, she probably will come back to the office again. However, it is likely for her to go home when this transition takes place at the time period 18:00 - 20:00. With the two temporal factors, NEXT calculates the hidden intent $\mathbf{h}^q_{t_{i}}$ as follows:
\begin{align}
\mathbf{h}^q_{t_i}&=ReLU\left(\mathbf{W}_\pi(t_i-t_{i-1})\left(\alpha\mathbf{q}^u_{t_{i-1}}+(1-\alpha)\mathbf{m}_{q^u_{t_{i-1}}}\right)+\mathbf{b}_{t_i}\right)\label{eqn:temp}
\end{align}
where $\mathbf{b}_{t_i}$ is the bias vector of the time slot within which $t_i$ falls. Here, the interval dependent transition in Equation~\ref{eqn:Wpi} is similar to the work in STRNN~\cite{aaai16:liu}. However, STRNN takes all historical POIs within the interval window for consideration in a recurrent manner, which is computational expensive. Further, STRNN does not consider  time-specific bias vector $\mathbf{b}_{t_i}$.

\begin{figure}{\includegraphics[width=.65\linewidth] {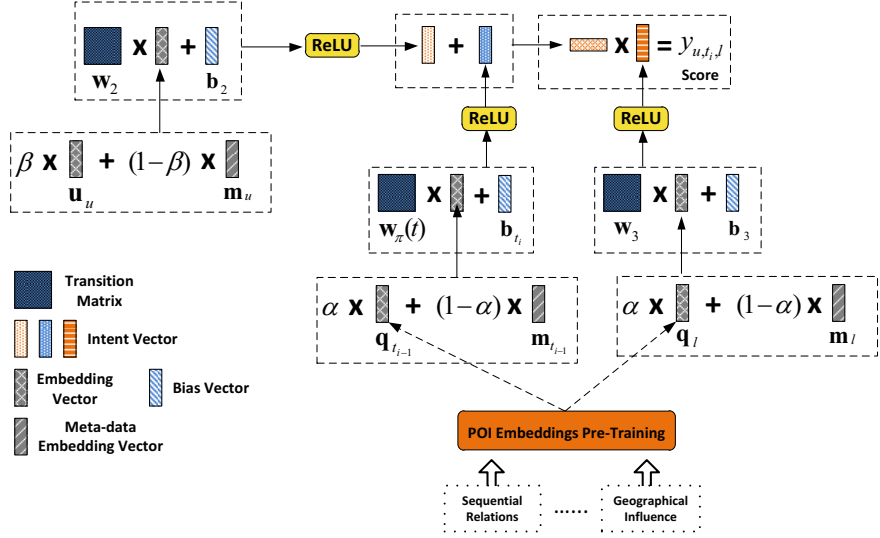}
 }%
 \centering
\caption{Overall Architecture of NEXT, $\mathbf{m}_u$, $\mathbf{m}_\ell$ and $\mathbf{m}_{q_{t_{i-1}}}$ are the embedding vectors of the meta-data associated with the user, candidate POI and the last visited POI respectively.}
\label{fig:framework}
\end{figure}

\subsection{POI Embeddings Pre-Training}\label{ssec:pretrain}
The sequential relations refer to the transition probability that a user visits POI $q_b$ after visiting POI $q_a$ (\ie $q_a\rightarrow q_b$). Hence, the transition probabilities convey the general transition patterns, (\eg from an airport to a hotel). Also, since users like to visit the nearby POIs and their activities are often constrained into a few regions, the visiting behaviors are affected a lot by the geographical influence.  Sequential relations and geographical influence are validated to be effective for the POI recommendation in many studies~\cite{kdd11:cho,sigir11:ye,sigir15:xutao,sigir15:zhang,ijcai15:feng}.

In NEXT, we propose a POI embedding pre-training strategy to encode the sequential relations and geographical influence among POIs. Because the non-convexity of the objective function in NEXT, there does not exist a global optimal solution. In such case, current optimization strategy is to find a local optimum. It is widely accepted that a good embedding initialization scheme could result in a faster convergence and superior performance of neural network models~\cite{www17:he}. In this sense, POI embedding pre-training can also benefit the model learning.

We adopt DeepWalk~\cite{kdd14:perozzi}, a network representation learning technique, to learn the embedding of each POI. DeepWalk builds short sequences of nodes based on random walk over the network structure. Then a neural language model SkipGram~\cite{arxiv13:mikolov} is adopted to learn the embeddings of the nodes by maximizing the probability of a node's neighbor in the sequences.

In order to retain these two kinds of information in the latent embedding space, we build a network structure by taking each POI as a distinct node in the network. Specifically, we create the random walk sequences over POIs by using a mixture of both the POI transition patterns and the geographical influence. The random walk transition from POI $q_i$ to POI $q_j$ over the network is calculated as follows:
\begin{align}
p(q_j|q_i)&=\rho\frac{\kappa(q_i,q_j)}{\sum_k   \kappa(q_i,q_k)}+(1-\rho)\frac{f_{q_i,q_j}}{\sum_k f_{q_i,q_k}}\label{eqn:randomwalk}\\
\kappa(q_i,q_j)&=1/(1+e^{5\frac{d(q_i,q_j)-\bar{d}}{\sigma(d)}})\label{eqn:dist}
\end{align}
where $d(q_i,q_j)$ denotes the Euclidean distance between POIs $q_i$ and $q_j$ by using their coordinates, $\bar{d}$ and $\sigma(d)$ are the mean and standard deviation of $d(q_i,q_j)$ respectively, $f_{q_i,q_j}$ is the transition frequency from $q_i$ to $q_j$ in the training dataset.

In Equation~\ref{eqn:randomwalk}, the first term in the right part captures the inherent geographical influence between POIs, while the second term captures the transition behaviors of massive users. $\rho$ is used here to balance the two components. For each POI, we generate $\tau$ random walks of length $r$ according to Equation~\ref{eqn:randomwalk} as in~\cite{kdd14:perozzi}. Then SkipGram language model with hierarchical softmax is applied over these random walk sequences. A POI's embedding is learnt to maximize the probability of seeing its neighbors in the sequences. Based on Equation~\ref{eqn:randomwalk}, the POIs that are close in geographical distance and likely to be visited successively by  users will be closer in the  embedding space. After finishing the embedding learning by SkipGram, we use the pre-trained POI embeddings as the initialization in model training. In the evaluation (Section~\ref{sec:exp}), we find that this pre-training strategy delivers better recommendation performance. The overall network architecture of NEXT is illustrated in Figure~\ref{fig:framework}.

\eat{
It was shown that the geographical proximity plays a major role for a user's next move~\cite{kdd11:cho,sigir15:xutao,sigir15:zhang,ijcai15:feng,aaai16:liu}. In this sense, sequential relations carry both the geographical information and POI transition patterns (\ie POI correlations).
}

Furthermore, we use the pre-trained POI embeddings to initialize user embedding $\mathbf{u}_u$. We first count the frequency of the POI a user $u$ has visited in the training dataset, and then use the normalized frequency as the weight to calculate the initial user embedding:
\begin{align}
\mathbf{u}_u&=\frac{1}{|\mathbf{L}_u|} \sum_j f^u_j\cdot \mathbf{q}_j
\end{align}
where $|\mathbf{L}_u|$ is the number of POI visits of user $u$ in the training set, $f^u_j$ is the frequency of POI $q_j$ being visited by user $u$.


\subsection{Cold-Start and Interpretation }\label{ssec:discuss}
\paratitle{Cold-Start.} The proposed NEXT can inherently handle POI recommendation for both cold-start users and cold-start POIs. In Equation~\ref{eqn:score}, the final intent calculation is the sum of $\mathbf{h}^q_{t_i}$ and $\mathbf{h}^u_{t_i}$. This additive mechanism has a potential merit for  cold-start problems. Given a new user with very few historical visits (\eg a single POI visit available), we can directly recommend the POIs based on Equation~\ref{eqn:score} by using $\mathbf{h}^q_{t_i}$ alone. Further, with Equation~\ref{eqn:intentusermeta}, we can calculate $\mathbf{h}^u_{t_i}$ by using her meta-data information $\mathcal{A}_u$ (\ie by setting $\beta=0$). This is particularly helpful for freshers that have no historical visit record. We will investigate the effectiveness of NEXT for cold-start users in Section~\ref{ssec:coldstart}.

For a cold-start POI $q$ that has not been visited by any user. It is possible to calculate $\mathbf{h}^q_{t_i}$ in Equation~\ref{eqn:intentpoimeta} based on its nearby POIs and meta-data information $\mathcal{A}_q$.

\eat{
Similarly, for a cold-start POI $q$ that have not been visited by any user, we can calculate $\mathbf{h}^q_{t_i}$ in Equation~\ref{eqn:intentpoimeta} based on its nearby POIs and the meta-data information $\mathcal{A}_q$. Specifically, given a new POI $q$, we can approximate its embedding $\mathbf{q}_q$ as follows:
\begin{align}
\mathbf{q}_q=\frac{1}{\sum_{q'\in\mathcal{N}_K(q)}s(q,q')}\sum_{q'\in \mathcal{N}_K(q)}s(q,q')\mathbf{q}_{q'}\label{eqn:coldstartpoi}
\end{align}
where $\mathcal{N}_K(q)$ is the set containing top-$K$ nearby POIs of $q$, $s(q,q')$ is the similarity score between POI $q$ and POI $q'$. In Equation~\ref{eqn:coldstartpoi}, we can estimate $s(q,q')$ based on the geographical distance, the similarity in terms of the meta-data information or their combination.
}

\paratitle{Interpretation.} Recall that the hidden intent calculations in Equations~\ref{eqn:candidatemeta},~\ref{eqn:intentpoimeta} and~\ref{eqn:intentusermeta} use the rectified linear unit (ReLU) as the nonlinear activation function. Given ReLU generates non-negative and sparse hidden vectors, it facilitates the interpretation for each individual hidden intent dimension. For example, consider the case that word description is associated with each POI. That is, the items in $\mathcal{A}_q$ are the words used to describe POI $q$. We can get the contribution vector $\vec{\omega}_w$ of word $w$ by setting $\alpha=0$ in Equation~\ref{eqn:candidatemeta}.
\begin{align}
\vec{\omega}_w=ReLU(\mathbf{W}_3\mathbf{m}_w+\mathbf{b}_3)\label{eqn:omega}
\end{align}
where $\mathbf{m}_w$ is the embedding vector of word $w$. Following a topical keyword re-ranking method proposed in~\cite{cikm09:song}, we assign a score for each word under a dimension as follows:
\begin{align}
\kappa_i(w)=\frac{\vec{\omega}_w(i)}{\sum_j \vec{\omega}_w(j)}\label{eqn:kappa}
\end{align}
where score $\kappa_i(w)$ reflects the preference of word $w$ under hidden dimension $i$. By examining the top words for each dimension in terms of $\kappa_i(w)$, we can obtain the semantic meaning of an intent dimension. This interpretability sheds light on many new enhancements for recommendation. For example, we can allow a user to adjust the recommendation system by explicitly highlighting her preferred intent dimensions. In this way, we can add an additional bias vector in Equation~\ref{eqn:score} to enhance this preference: $y_{u,t_i,l}=(\mathbf{h}^{u}+\mathbf{s}_u+\mathbf{h}_{t_i}^q)^T\mathbf{c}_l$, where $\mathbf{s}_u$ is  user specified intent preference. Experimental studies are presented in Section~\ref{ssec:interpretation}.

\subsection{Training}\label{ssec:training}
The parameters of our model are: $\Theta=\{\mathbf{W}_{\ast},\mathbf{M},\mathbf{B},\mathbf{U},\mathbf{Q},\mathbf{b}_2,\mathbf{b}_3\}$, where $\mathbf{W}_{\ast}$ refers to all transition matrices $\mathbf{W}_0$, $\mathbf{W}_2$, $\mathbf{W}_3$, $\mathbf{W}_{\pi}$; and $\mathbf{M}$ contains all item embeddings for the associated meta-data of both users and POIs; $\mathbf{B}$ contains all time slot bias vectors $\mathbf{b}_t$; $\mathbf{U}$ contains all user embeddings, and $\mathbf{Q}$ contains all POI embeddings.

The model training aims to optimize above parameters such that each POI visit in the sequence of a user's POI visits in the training set can be predicted successfully. We adopt a softmax function to calculate the predicted POI probability vector $\mathbf{p}^u_{t_i}$ for user $u$ at time $t_i$:
\begin{equation}
\mathbf{p}^u_{t_i}(k)=\frac{e^{y_{u,t_{i},k}}}{\sum_j e^{y_{u,t_{i},j}}}\label{eqn:softmax}
\end{equation}
Then, we use the cross-entropy error between the ground truth POI distribution (\ie in a one-hot form) and predicted POI distribution by Equation~\ref{eqn:softmax} as the cost objective:
\begin{equation}
J=\frac{1}{U}\sum^U_u\sum_{t\in \mathbf{L}_u}\sum_k^Q \hat{\mathbf{q}}^u_{t}(k)\cdot\log\mathbf{p}^u_{t}(k)+\lambda\|\Theta\|_2\label{eqn:objective}
\end{equation}
where $\mathbf{L}_u$ is the set of historical POI visits in the training set for user $u$, $Q$ is the number of all POIs under consideration, $\hat{\mathbf{q}}^u_{t}$ is the ground truth POI distribution at time $t$ with \textit{1-of-Q} coding scheme, $\lambda$ controls the importance of the regularization term, and $U$ is the number of users under consideration.

To minimize the objective, we use stochastic gradient descent (SGD)~\cite{nn91:bottou} and back propagation to update the parameters. Although POI embeddings $\mathbf{Q}$ is pre-trained based on the sequential relations and geographical influence, we further fine-tune the embeddings based on the cost objective.

\section{Experiments}\label{sec:exp}
In this section, we conduct  experiments to evaluate the proposed NEXT against the state-of-the-art alternatives over three real-world datasets.

\subsection{Datasets}\label{ssec:datasets}

\paratitle{Foursquare Singapore (SIN)} dataset is a collection of $194,108$ check-ins made within Singapore from $2,321$ users at $5,596$ POIs between Aug.~2010 and Jul.~2011 in Foursquare~\cite{sigir13:yuan}. This dataset has previously been used in other studies~\cite{sigir13:yuan,ijcai15:feng,sigir15:xutao}.

\paratitle{Gowalla} dataset contains $736,148$ check-ins made within California and Nevada between Feb.~2009 and Oct.~2010 in Gowalla~\cite{kdd11:cho}. The \textit{Gowalla} dataset has previously been used in~\cite{sigir13:yuan,ijcai15:feng,sigir15:xutao,aaai16:liu}.

\paratitle{CA} dataset is a collection of $483,813$ check-ins made in Foursquare by $4,163$ users living in California. Each distinct POI is provided with a text description indicating its content. There are total $50$ distinct words in all descriptions. Moreover, each user is connected to a number of other users (\ie friendship). This dataset has previously been used in~\cite{tois16:yin}. Note that, this is the  only dataset that contains auxiliary meta-data for both users and POIs.

In all three datasets, each check-in is associated with a timestamp indicating when the user made this check-in. Following the work of PRME-G in~\cite{ijcai15:feng}, we remove the less frequent users and POIs from each dataset, such that each user has at least $10$ check-ins, and each POI has been visited by at least $10$ users. The data statistics on these three datasets after preprocessing is reported in Table~\ref{tbl:datasets}. In \textit{CA} dataset, there are on average $2.67$ descriptive words for a POI and $4.36$ friends for a user.

\begin{table}[t]
\small
\centering
\caption{Statistics on the three datasets. \#User: the number of users; \#POI: the total number of POIs; \#Check-in: the total number of check-ins; \#AvgC: average number of check-ins per user; \#Avg($\mathcal{A}_u$): average number of items in $\mathcal{A}_u$; \#Avg($\mathcal{A}_q$): average number of items in $\mathcal{A}_q$.}
\label{tbl:datasets}
\begin{tabular}{@{}c|crrrcc@{}}
\toprule
& & & & & \multicolumn{2}{c}{Meta-data}\\
\cline{6-7} \raisebox{1.5ex}[0pt]{Dataset} & \raisebox{1.5ex}[0pt]{\#User} & \raisebox{1.5ex}[0pt]{\#POI} & \raisebox{1.5ex}[0pt]{\#Check-in} & \raisebox{1.5ex}[0pt]{\#AvgC} & \#Avg($\mathcal{A}_u$)& \#Avg($\mathcal{A}_q$)\\\midrule
SIN & 1,918 & 2,678 & 155,514 & 81.08 & - & -\\ 
Gowalla & 5,073 & 7,020 & 252,945 & 49.86 & - & - \\
CA & 2,031 & 3,112 & 105,836 & 52.1 & 4.36 & 2.67\\\bottomrule
\end{tabular}
\end{table}

\begin{table}
\scriptsize
\centering
\caption{Performance comparison over three datasets by Acc@K and MAP. The best results are highlighted in boldface on each dataset. $\dagger$ indicates that the difference to the best result is statistically significant at $0.05$ level.}
\label{tbl:cmp}
\begin{tabular}{@{}c||ccc|c||ccc|c||ccc|c@{}}
\toprule
& \multicolumn{4}{c||}{SIN} & \multicolumn{4}{c||}{Gowalla}& \multicolumn{4}{c}{CA}\\
\cline{2-13}\raisebox{1.5ex}[0pt]{Method} & Acc@1 & Acc@5 & Acc@10 & MAP & Acc@1 & Acc@5 & Acc@10 & MAP & Acc@1 & Acc@5 & Acc@10 & MAP\\
\midrule
 PMF & $0.0013^\dagger$ & $0.0311^\dagger$ & $0.0731^\dagger$ & $0.0235^\dagger$ & $0.0002^\dagger$ & $0.0149^\dagger$ & $0.0418^\dagger$ & $0.0125^\dagger$ & $0.0006^\dagger$ & $0.0050^\dagger$ & $0.0109^\dagger$ & $0.0106^\dagger$ \\
 PRME-G & $0.0751^\dagger$ & $0.1156^\dagger$ & $0.1357^\dagger$ & $0.0991^\dagger$ & $0.1088^\dagger$ & $0.1600^\dagger$ & $0.1783^\dagger$ & $0.1348^\dagger$ & $0.0888^\dagger$ & $0.1287^\dagger$ & $0.1520^\dagger$ & $0.1130^\dagger$\\
 Rank-GeoFM & $0.0705^\dagger$ & $0.1870^\dagger$ & $0.2575^\dagger$ & $0.1313^\dagger$ & $0.0488^\dagger$ & $0.1428^\dagger$ & $0.1997^\dagger$ & $0.1000^\dagger$ & $0.0540^\dagger$ & $0.1505^\dagger$ & $0.2085^\dagger$ & $0.1061^\dagger$\\
 GE & $0.0123^\dagger$ & $0.0486^\dagger$ & $0.0735^\dagger$ & $0.0326^\dagger$ & $0.0100^\dagger$ & $0.0158^\dagger$ & $0.0488^\dagger$ & $0.0281^\dagger$& $0.0894^\dagger$ & $0.1402^\dagger$ & $0.1651^\dagger$ & $0.1174^\dagger$\\\midrule
 NeuMF & $0.025^\dagger$ & $0.0854^\dagger$ & $0.1341^\dagger$ & $0.0654^\dagger$ & $0.0230^\dagger$ & $0.0682^\dagger$ & $0.1082^\dagger$ & $0.0549^\dagger$& $0.0437^\dagger$ & $0.0944^\dagger$ & $0.1361^\dagger$ & $0.0781^\dagger$\\
 RNN & $0.1063^\dagger$ & $0.2397^\dagger$ & $0.3072^\dagger$ & $0.1742^\dagger$ & $0.084^\dagger$ & $0.1859^\dagger$ & $0.2364^\dagger$ & $0.1376^\dagger$& $0.0865^\dagger$ & $0.1877^\dagger$ & $0.2370^\dagger$ & $0.1397^\dagger$\\
 LSTM & $0.1032^\dagger$ & $0.2344^\dagger$ & $0.3015^\dagger$ & $0.1701^\dagger$ & $0.0868^\dagger$ & $0.1979^\dagger$ & $0.2535^\dagger$ & $0.1443^\dagger$& $0.0931^\dagger$ & $0.2028^\dagger$ & $0.2583^\dagger$ & $0.1511^\dagger$\\
 GRU & $0.0999^\dagger$ & $0.2211^\dagger$ & $0.2864^\dagger$ & $0.1626^\dagger$ & $0.0838^\dagger$ & $0.2015^\dagger$ & $0.2644^\dagger$ & $0.1454^\dagger$& $0.0924^\dagger$ & $0.1974^\dagger$ & $0.2505^\dagger$ & $0.1482^\dagger$\\
 STRNN & $0.0826^\dagger$ & $0.1948^\dagger$ & $0.2636^\dagger$ & $0.1431^\dagger$ & $0.0557^\dagger$ & $0.1539^\dagger$ & $0.2081^\dagger$ & $0.1079^\dagger$& $0.0713^\dagger$ & $0.1637^\dagger$ & $0.2181^\dagger$ & $0.1221^\dagger$\\\midrule
 NEXT & \textbf{0.1358} & \textbf{0.2897} & \textbf{0.3673} & \textbf{0.2127} & \textbf{0.1282} & \textbf{0.2644} & \textbf{0.3339} & \textbf{0.1975}& \textbf{0.1115} & \textbf{0.2396} & \textbf{0.3038} & \textbf{0.1772}\\\bottomrule
\end{tabular}
\end{table}

\subsection{Experimental Setup}\label{ssec:setup}

\paratitle{Methods and parameter settings.} We compare our model against the following recent state-of-the-art POI recommendation approaches.

\begin{itemize}
\item \textbf{PMF} is a method based on conventional probabilistic matrix factorization over the user-POI matrix~\cite{nips07:ruslan}.

\item \textbf{PRME-G} embeds user and POI into the same latent space to capture the user transition patterns~\cite{ijcai15:feng}. The geographical influence is incorporated in PRME-G through a simple weighing scheme. We use the recommended settings with $60$ dimensions and $\pi=6h$ as in their paper.

\item \textbf{Rank-GeoFM} is a ranking based geographical factorization approach~\cite{sigir15:xutao}. Rank-GeoFM learns the embeddings of users, POIs by fitting the user's POI frequency. Both temporal context and geographical influence are incorporated in a weighting scheme. We use the recommended settings with $K=100$, $k=300$ as in their paper and fine-tune the parameters $\alpha$ and $\beta$ on the development set.

\item \textbf{Graph based Embedding (GE)} jointly learns the embeddings of POIs, regions, time slots, and auxiliary meta-data (\ie descriptive words of POIs) in one common hidden space~\cite{cikm16:xie}. The recommendation score is then calculated by a linear combination of the inner products for these contextual factors. We tune hyper-parameters $N$ and $\triangle T$ on the development set.

\item \textbf{Neural Matrix Factorization (NeuMF)} is a recent state-of-the-art deep neural network based algorithm over implicit feedback~\cite{www17:he}. NeuMF combines both generalized matrix factorization and MLP under one framework to learn latent features. Like PMF, we apply NeuMF over the user-POI matrix for the recommendation. The best performance is reported by tuning hyper-parameters.

\item \textbf{STRNN} is a RNN-based model for next POI recommendation~\cite{aaai16:liu}. It incorporates both the temporal context and geographical information within recurrent architecture.

\item \textbf{RNN} is a standard RNN model for sequence modeling, upon which the above STRNN model was built~\cite{interspeech10:mikolov}. In the context of POI recommendation, the hidden feature vector $\mathbf{h}^u_{t_{i}}$ of user $u$ at time $t_{i}$ is calculated recurrently based on the whole historical POI visits:
    \begin{equation}
    \mathbf{h}^u_{t_i}=\sigma(\mathbf{W_4}\mathbf{q}^u_{t_{i-1}}+\mathbf{C}\mathbf{h}^u_{t_{i-1}})
    \end{equation}
    where $\mathbf{W_4}$ is the transition matrix from the input embedding to the hidden state, $\mathbf{C}$ is the state-to-state recurrent weight matrix, $\sigma$ is chosen to be the sigmoid function. Following the work in~\cite{aaai16:liu}, we calculate the recommendation score $y_{u,t_{i},\ell}$ of POI $\ell$ for user $u$ at time $t_{i}$  as follows:
    \begin{equation}
    y_{u,t_i,\ell}=(\mathbf{h}^u_{t_i}+\mathbf{u}_u)^T\mathbf{q}_{\ell}\label{eqn:rnnscore}
    \end{equation}

\item \textbf{LSTM} is an variant of RNN model which contains a memory cell and three multiplicative gates to allow long-term dependency learning~\cite{nc97:sepp}. We calculate the recommendation score by using Equation~\ref{eqn:rnnscore}.

\item \textbf{GRU} is a variant of RNN model which is equipped with two gates to control the information flow~\cite{corr14:cho}. We calculate the recommendation score by using Equation~\ref{eqn:rnnscore}.
\end{itemize}

Other possible alternatives are empirically found to be inferior to STRNN, PRME-G, and Rank-GeoFM, in their works respectively\footnote{Some recent works (\eg \cite{aaai16:he,aaai16:zhao}) that incorporate POI categories and date information, are excluded for comparison, because our datasets do not contain these meta-data.}. Hence, due to space limitation, we leave these comparisons to our future work. Also, the proposed TRM model in~\cite{tois16:yin} can be evaluated based on \textit{CA} dataset. However, due to the shortness of POI description and smaller number of POIs after preprocessing, TRM only achieves a slightly better performance than PMF. Therefore, we exclude TRM from further comparison. The first four comparative methods listed above are conventional matrix factorization or embedding learning based techniques. The next five methods are Neural Networks based methods, which apply the nonlinearity for high-level transformation. Note that GRU and LSTM have not been evaluated in previous work on next POI recommendation task. For performance evaluation, we use the last $20\%$ POI visits of each user as test set, the earliest $70\%$ POI visits as training set, and the remaining $10\%$ data as validation set to tune parameters.

\paratitle{Metrics.} Following the existing works~\cite{cikm16:xie,aaai16:liu,aaai16:he}, two standard metrics are used for performance evaluation: Acc@K and Mean Average Precision (MAP). For a specific test instance (\ie a user visited a POI in the test set), Acc@K is $1$ if the visited POI appears in the top-K ranking; otherwise $0$ is taken. The overall Acc@K is the average value over all test instances. Here, we choose to report Acc@K with $K=\{1, 5, 10\}$. MAP is widely used to evaluate the quality of ranking. The higher the ground truth POI is ranked, the larger the MAP value, which indicates a better performance.

\paratitle{Hyperparameters and Training.} The interval threshold $\pi$ in Equation~\ref{eqn:Wpi} is empirically set to be 6/6/72 hours for \textit{SIN}, \textit{Gowalla} and \textit{CA} datasets respectively. The dimensionality for the embeddings and the hidden intent are fixed to be $60$ for neural network based methods for fair comparison (\ie $d=60$ in NEXT). The regularization parameter $\lambda$ is $0.01$ and the learning rate $\gamma$ is $0.005$. As to incorporating auxiliary meta-data information, we set $\alpha=0.3,\beta=0.2$ in NEXT. We apply the early stop based on the validation set, or a maximum of $50$ epochs are run for neural network based methods.

As to POI embeddings pre-training, we set $\tau=50$ and $r=20$ as in the original work of DeepWalk~\cite{kdd14:perozzi}. In Equation~\ref{eqn:randomwalk}, $\rho=0$ is used in generating random walks for the performance comparison. The impact of $\rho$ will be studied in Section~\ref{ssec:analysis}.

\subsection{Performance Comparison}\label{ssec:comparison}
For performance comparison, we report the recommendation accuracy  of different methods over the three datasets in Table~\ref{tbl:cmp}, where significance test is by  Wilcoxon signed-rank test. We make the following observations:

First, the proposed NEXT model performs significantly better than all existing state-of-the-art alternatives evaluated here on the three datasets in all the metrics. Specifically, NEXT outperforms the conventional matrix factorization method PMF significantly by a large margin. As to the three embedding learning based solutions (\ie PRME-G, Rank-GeoFM, GE), NEXT outperforms them by around $62.0\%$ - $552.5\%$, $46.5\%$ - $602.8\%$ and $50.9\%$ - $67.0\%$ in terms of MAP metric on \textit{SIN}, \textit{Gowalla} and \textit{CA} datasets respectively. Note that both PRME-G and Rank-GeoFM incorporate information from temporal context and geographical influence within their models on \textit{SIN} and \textit{Gowalla}. The large improvement suggests that high-level intent features extracted through a nonlinearity in NEXT can better catch the user's spatial behaviors. Moreover, NEXT consistently outperforms four RNN-based methods: RNN, LSTM, GRU, and STRNN. The performance gain provided by NEXT over these four counterparts is about $22.1\%$ - $48.6\%$ and $35.8\%$ - $83.0\%$ in terms of MAP metric on the \textit{SIN} and  \textit{Gowalla} respectively. This indicates that the mechanism to absorb two kinds of temporal context in NEXT is effective for the task of next POI recommendation.

Second, PMF performs the worst on three datasets in all metrics, because the user-POI matrix is very sparse on these datasets, and no temporal context or geographical influence is leveraged at all. Similar results are observed on NeuMF, a neural network based collaborative filtering technique based on implicit feedback information. Since both PRME-G and Rank-GeoFM utilize ranking based optimization strategy, the data sparsity issue is alleviated by making use of unobserved data to learn the parameters. Moreover, temporal information and geographical influence are incorporated in these two models. Therefore, a large performance improvement is obtained by PRME-G and Rank-GeoFM over PMF and NeuMF. The same phenomenon was also observed in the related works~\cite{sigir15:xutao,ijcai15:feng,aaai16:liu}.

Third, NeuMF significantly outperforms conventional PMF. This suggests the superiority of nonlinearity for extracting hidden high-level features. As being an embedding learning technique, GE performs much worse than PRME-G and Rank-GeoFM on both \textit{SIN} and \textit{Gowalla} datasets. This is reasonable because no region information is available on these two datasets. The region information works as the geographical influence for GE model. However, region information is provided in \textit{CA} dataset;  we observe that GE achieves very close performance to PRME-G and Rank-GeoFM.

Fourth, the three RNN-based methods (\ie RNN, LSTM, GRU) perform much better than PMF, PRME-G and Rank-GeoFM in most metrics. This is consistent with our above discussion that the non-linear transformation operation as provided by the neural network models enables better high-level spatial intent learning. Although LSTM and GRU were designed to alleviate the \textit{exploding or vanishing gradients} problem, no superiority is observed for them over RNN model on the \textit{SIN} dataset. Reported in Table~\ref{tbl:datasets}, the users in \textit{SIN} have more POI visits on average. Because RNN-based models accumulate all historical information in the last hidden feature vector~\cite{acl16:wang}, the longer POI sequence could introduce much irrelevant information that hurts the performance. This result indicates that the visiting behaviors performed a long time ago are irrelevant for next POI recommendation. Also, we observe that STRNN only achieves close performance with Rank-GeoFM and PRME-G.

In summary, the experimental results show that the proposed NEXT can successfully learn  user's spatial intent, leading to superior performance of next POI recommendation.

\begin{figure}
\centering
\begin{subfigure}[MAP]{\includegraphics[width=.42\columnwidth] {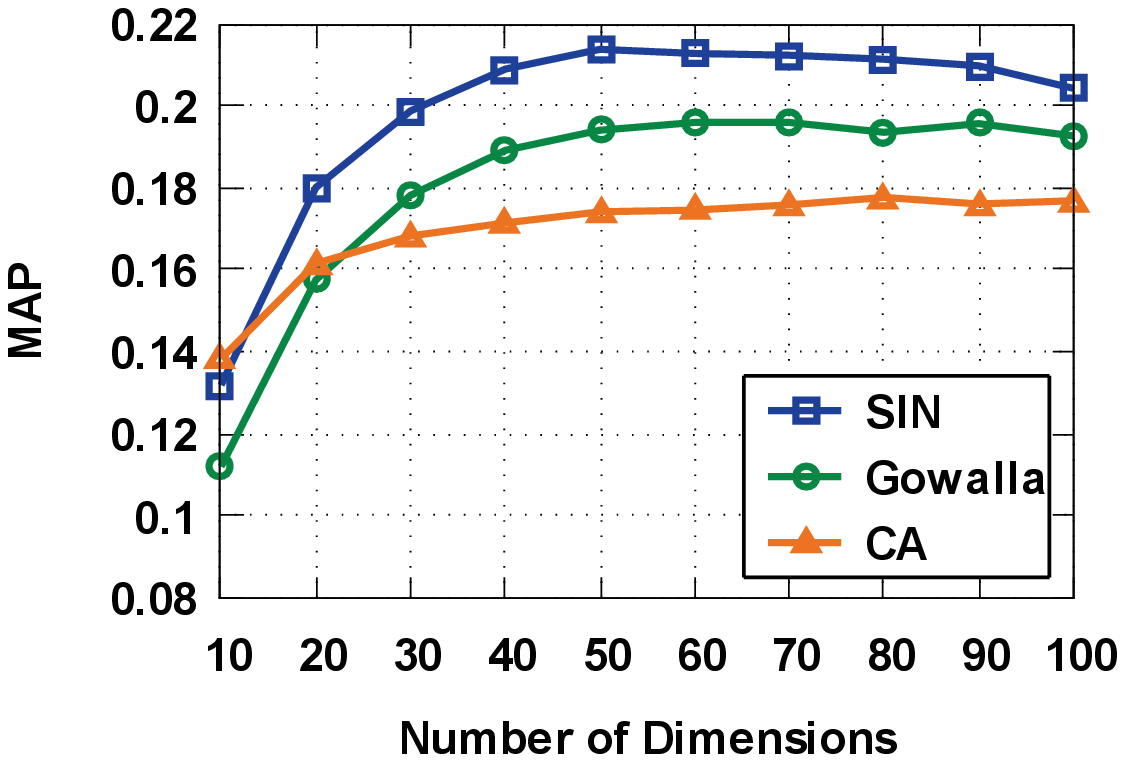}
   \label{fig:MAP}
 }%
\end{subfigure}\hfill
 \begin{subfigure}[Acc@10]{\includegraphics[width=.42\columnwidth]{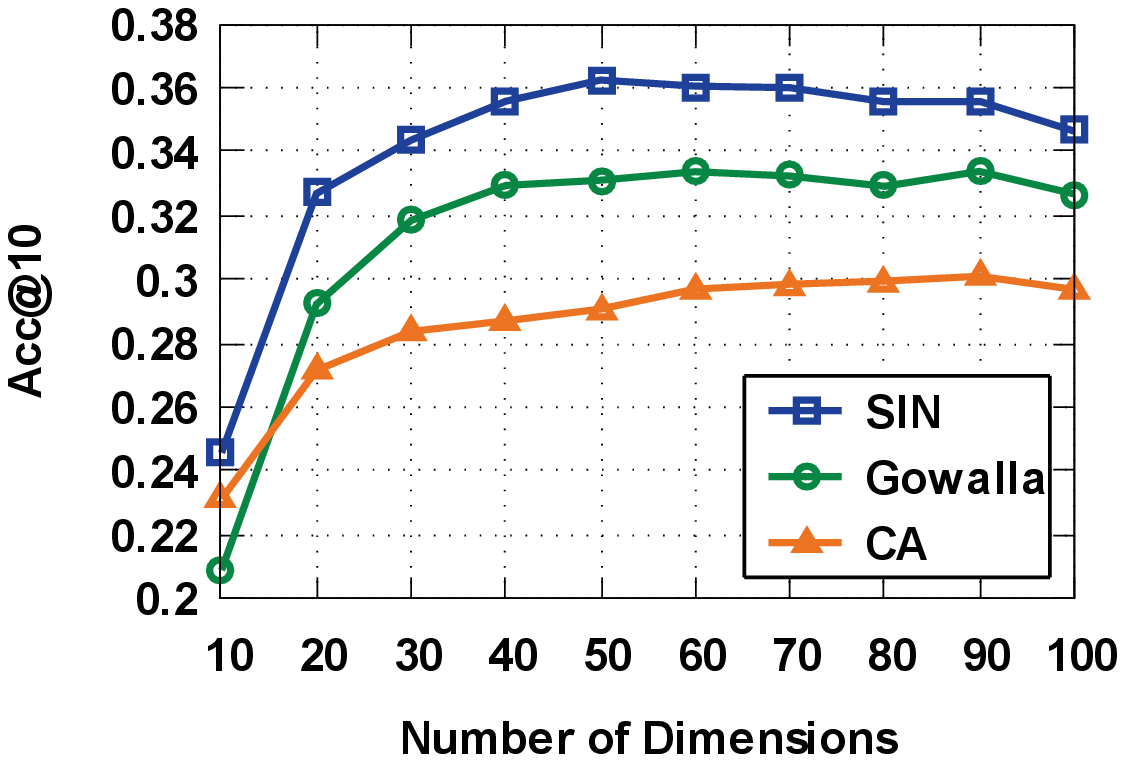}
   \label{fig:recall}
 }%
\end{subfigure}
\caption{Effect of the number of dimensions in NEXT}
\label{fig:dimension}
\end{figure}

\subsection{Experiments on Cold-Start }\label{ssec:coldstart}
Here, we evaluate the performance of NEXT and other competitors for cold-start users. Specifically, since each dataset is preprocessed to retain only active users and POIs (ref. Section~\ref{ssec:datasets}), we therefore take $200$ inactive users that were excluded from the training for evaluation. We conduct the experiments on  \textit{CA} dataset, since it is the only dataset containing auxiliary meta-data information.

For each cold-start user $u$, we randomly pick a POI transition record $(q_i,q_j)$ such that the user visited $q_j$ after her latest visit at $q_i$. For evaluation purpose, we restrict to the record of both $q_i$ and $q_j$ being included in the training set. Here, we test to recommend $q_j$ by utilize both her latest POI visit and meta-data. Among the baseline methods, only PRME-G, STRNN, RNN, LSTM and GRU can be adapted here by utilizing only the POI information. STRNN, RNN, LSTM and GRU are all RNN-based models. Since LSTM achieves the best performance on \textit{CA} dataset among these RNN variants (ref. Table~\ref{tbl:cmp}), we choose LSTM as the representative, and report its performance for cold-start user recommendation. Other variants are found to be inferior than LSTM for this experiment. Table~\ref{tbl:colduser} reports the performance of different methods. We observe that NEXT outperforms PRME-G and LSTM in most metrics. This suggests that incorporating meta-data information is positive for addressing the recommendation for cold-start users.

\eat{
\paratitle{Cold-Start POIs.} For each cold-start POI $q_c$, we randomly pick a user that has a POI transition record $(q_i,q_c)$ such that $q_i$ is included in the training set. Here, we test to recommend $q_c$ in three scenarios: (1) utilize top-$K$ neighbors without considering geographical distance, \ie setting $s(q_c,q_n)=1$ in Equation~\ref{eqn:coldstartpoi}) (NoDist); (2) utilize geographical distance for neighboring POI weighting, \ie setting $s(q_c,q_n)=\kappa(q_c,q_n)$ in Equation~\ref{eqn:coldstartpoi} (Dist); (3) utilize the similarity between $q_c$ and its neighbor $q_n$, \ie $s(q_c,q_n)=sim(\mathcal{A}_{q_c},\mathcal{A}_{q_n})$ in Equation~\ref{eqn:coldstartpoi}. We adopt the cosine similarity with TF-IDF scheme to calculate $sim(\mathcal{A}_{q_c},\mathcal{A}_{q_n})$ based on the textual descriptions. Among the baseline methods, only GE can be adapted to address. Here, we the learning process for the embedding of $q_c$ in GE is conducted by fixed all the embeddings learnt after the model training. For the other RNN based methods, we conduct the recommendation based
}

\begin{table}
\centering
\caption{Performance comparison for cold-start users.}
\label{tbl:colduser}
\begin{tabular}{c||ccc|c}
\toprule
 Method & Acc@1 & Acc@5 & Acc@10 & MAP \\
\midrule
  PRME-G & $0.0550$ & $0.0650$ & $0.0800$ & $0.0631$ \\
  LSTM  & $0.0300$ & $0.1200$ & $\textbf{0.1900}$ & $0.0765$\\
  NEXT  & $\textbf{0.0600}$ & $\textbf{0.1400}$ & $0.1850$ & $\textbf{0.1045}$\\
 \bottomrule
\end{tabular}
\end{table}

\eat{
\begin{table}
\scriptsize
\centering
\caption{Performance comparison for cold-start users and POIs.}
\label{tbl:colduser}
\begin{tabular}{|c|c|c||c|c|c|c|}
\toprule
Dataset & Method & Mode & Acc@1 & Acc@5 & Acc@10 & MAP \\
\midrule
 \multirow{3}{*}{Gowalla} & PRME-G & $q_i$ & $xxx$ & xxx & xxx & xxx \\
 & LSTM & $q_i$ & $xxx$ & $xxx$ & $xxx$ & $xxx$\\
 & NEXT & $q_i$ & $0.0200$ & $0.0600$ & $0.0800$ & $0.0474$\\
\bottomrule
 \multirow{5}{*}{CA} & PRME-G & $q_i$ & $xxx$ & xxx & xxx & xxx \\
 & LSTM & $q_i$ & $xxx$ & $xxx$ & $xxx$ & $xxx$\\
 & NEXT & $q_i$ & $0.0550$ & $0.1000$ & $0.1200$ & $0.0813$\\
 & NEXT & $\mathbf{m}_u$ & $0.0050$ & $0.0500$ & $0.0750$ & $0.0333$\\
 & NEXT & $q_i$+$\mathbf{m}_u$ & $0.0600$ & $0.1200$ & $0.2050$ & $0.1089$\\
 \bottomrule
\end{tabular}
\end{table}
}

\subsection{Interpretation}\label{ssec:interpretation}
Now, we evaluate the interpretability of NEXT based on the descriptive words associated with POIs on \textit{CA} dataset. We manually examine the top-$10$ words in terms of $\kappa_i(w)$ for each hidden dimension $i$. If these top words could convey a coherent and meaningful topic reflecting a person's activities, we consider a dimension as being interpretable. As the result, we find $34$ interpretable dimensions among the 60 dimensions. Note that there are only $50$ unique words used in \textit{CA} dataset, and the dimension number  (\ie $60$) is even larger than the number unique words. Hence, we consider this result to be excellent. To further demonstrate the superiority of NEXT in producing interpretable hidden dimensions, we list the top-$5$ words for $5$ interpretable dimensions learnt by NEXT in Table~\ref{tbl:topwords}. The top words in each dimension can be easily interpreted to cover a topic on a specific activity. For example, dimension 1 expresses the activity of enjoying nightlife by watching movies; dimension 3 talks about  outdoors recreation such as arts performing.

\begin{table}
\centering
\caption{Top-$5$ words of some interpretable dimensions by NEXT.}
\label{tbl:topwords}
\begin{tabular}{lllll}
\toprule
Dim 1 & Dim 2 & Dim 3 & Dim 4 & Dim 5 \\
\midrule
nightlife & travel & recreation & shop & park \\
spot & transport & outdoors & service & theme \\
food & airport & arts & hotel & venue \\
theater & hotel & entertainment & office & performing \\
movie & store & performing & clothing & drink \\
\bottomrule
\end{tabular}
\end{table}

\subsection{Analysis of NEXT}\label{ssec:analysis}
We now investigate the impact of different parameter settings in NEXT. Note that when studying a specific parameter, we set the other parameters to the values used in Section~\ref{ssec:setup}.

\paratitle{Temporal Context.} We first investigate the effect of the two kinds of temporal contexts in NEXT. Table~\ref{tbl:tc} lists the performance comparison over three datasets, where \checkmark refers to the model with the corresponding temporal context. Observe that incorporating either time interval or visit time information leads to better performance. More performance gain is obtained by introducing the time interval dependent transition, compared to using visit time specific preference alone. This validates that the time interval since the latest POI visit plays a critical role in learning spatial intent from historical spatial behavior. Further improvement is obtained by incorporating both time interval and visit time information together. This indicates that these two kinds of temporal context provide complementary benefits for next POI recommendation.

\eat{
We further study the impact of $\pi$ value in NEXT. Recall in Equation~\ref{eqn:Wpi}, a larger $\pi$ indicates that the user's spatial intent change temporally slower, while a smaller $\pi$ indicates that the user's spatial intent is mainly determined by the nearby movement. Tabel~xxx reports the performance of different $\pi$ values over the two datasets.
}

\begin{table}
\centering
\caption{Effect of the temporal context: time interval (TI) and time slot (TS).}
\label{tbl:tc}
\begin{tabular}{c|cc||ccc|c}
\toprule
Dataset & TI & TS & Acc@1 & Acc@5 & Acc@10 & MAP \\
\midrule
 \multirow{4}{*}{SIN} & - & - & 0.1161 & 0.2576 & 0.3250 & 0.1869 \\
 & \checkmark & - & 0.1322 & 0.2833 & 0.3569 & 0.2077 \\
 & - & \checkmark & 0.1272 & 0.2690 & 0.3414 & 0.1986 \\
 & \checkmark & \checkmark &\textbf{0.1358} & \textbf{0.2897} & \textbf{0.3673} & \textbf{0.2127}\\
\bottomrule
 \multirow{4}{*}{Gowalla} & - & - & 0.0986 & 0.2254 & 0.2861 & 0.1630 \\
 & \checkmark & - & 0.1172 & 0.2535 & 0.3250 & 0.1868 \\
 & - & \checkmark & 0.1058 & 0.2310 & 0.2919 & 0.1691 \\
 & \checkmark & \checkmark &\textbf{0.1282} & \textbf{0.2644} & \textbf{0.3339} & \textbf{0.1975}\\
\bottomrule
 \multirow{4}{*}{CA} & - & - & 0.0942 & 0.2104 & 0.2661 & 0.1553 \\
 & \checkmark & - & 0.0994 & 0.2185 & 0.2782 & 0.1607 \\
 & - & \checkmark & 0.1058 & 0.2281 & 0.2898 & 0.1691 \\
 & \checkmark & \checkmark &\textbf{0.1115} & \textbf{0.2396} & \textbf{0.3038} & \textbf{0.1772}\\
 \bottomrule
\end{tabular}
\end{table}

\paratitle{Number of Dimensions.} We study the effect of the number of dimensions of  hidden vectors and POI embeddings. Here, we vary the dimension number from $10$ to $100$. Figure~\ref{fig:dimension} shows the MAP and Acc@10 values for varying dimension numbers on the three datasets.  NEXT achieves stable performance in the range of $[50,100]$. We observe that NEXT  outperforms RNN, LSTM and GRU even when the number of dimensions is as small as $20$. The results further confirm the superiority of the proposed NEXT for next POI recommendation.

\paratitle{POI Embeddings Pre-training.} DeepWalk is used to generate POI sequences in NEXT to encode the sequential relations and geographical influence among POIs. The proportion parameter $\rho$ is used to balance the geographical influence and transition behavior components between two POIs (Equation~\ref{eqn:randomwalk}). Given the similar performance patterns observed for all three datasets, we choose to report the performance in \textit{CA} dataset only, due to space limitation.

Table~\ref{tbl:rho} reports the performance of different $\rho$ values over \textit{CA} dataset, where symbol $-$ refers to the model without using the pre-trained POI embeddings for initialization. First, we observe that the models initialized with pre-trained POI embeddings outperform the model without this initialization by a large margin. This validates the effectiveness of utilizing geographical distance and transition pattern between two POIs to pre-train POI embeddings. Second, all the settings with varying positive $\rho$ values achieve similar performance. And the best performance is achieved when $\rho=0$, \ie no geographical influence factor is exploited at all. This suggests that the geographical distance and transition patterns do not contain complementary information. Based on Tobler's first law of geography, ``Everything is related to everything else, but near things are more related than distant things." This indicates that when a user visits the next place, she will likely to visit a place near the place she visited from last time. In this sense, the geographical influence could be encoded within the transition patterns, as being validated by the results. Accordingly, we set $\rho=0$ in our experiments.

\begin{table}
\centering
\caption{Performance of NEXT on \textit{CA} with varying $\rho$ values}
\label{tbl:rho}
\begin{tabular}{c|ccc|c}
\toprule
& \multicolumn{4}{c}{CA} \\
\cline{2-5}\raisebox{1.5ex}[0pt]{$\rho$} & Acc@1 & Acc@5 & Acc@10 & MAP \\
\midrule
- & $0.0383$ & $0.101$ & $0.1406$ & $0.0744$ \\
 0 & \textbf{0.1115} & \textbf{0.2396} & \textbf{0.3038} & \textbf{0.1772} \\
0.3 & $0.1023$ & $0.2145$ & $0.2747$ & $0.1626$ \\
0.5 & $0.1015$ & $0.2077$ & $0.2713$ & $0.1605$ \\
0.7 & $0.1026$ & $0.2148$ & $0.2779$ & $0.1628$ \\
1 & $0.106$ & $0.221$ & $0.2813$ & $0.1666$ \\\bottomrule
\end{tabular}
\end{table}

\begin{figure}[t]
\centering
\includegraphics[width=0.35\columnwidth]{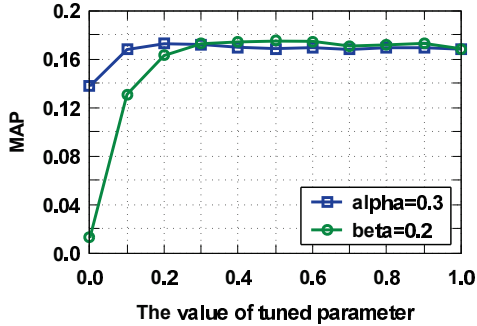}
\caption{Performance of NEXT with different $\beta$ and $\alpha$ values by fixing $\alpha=0.3$ and $\beta=0.2$ respectively.}
\label{fig:alphaBeta}
\end{figure}

\paratitle{Auxiliary Meta-data.} We further study the impact of incorporating auxiliary meta-data information to the recommendation accuracy in NEXT. Table~\ref{tbl:metadata} reports the performance with/without incorporating the associated friendship and textual description on \textit{CA}. We observe that NEXT achieves significant better performance by incorporating auxiliary meta-data information. Note that $\alpha$ and $\beta$ in Equations~\ref{eqn:temp},~\ref{eqn:candidatemeta} and~\ref{eqn:intentusermeta} control the importance of the meta-data of POIs and users respectively. Here, these two parameters are tuned in the following way. First, we choose the optimal $\alpha$ value by fixing $\beta=1.0$ (\ie no meta-data is used for uesr-based intent calculation). Then the optimal $\beta$ value is chosen by fixing this $\alpha$ value. Following this strategy, we set $\alpha=0.3$ and $\beta=0.2$ for \textit{CA} dataset. Figure~\ref{fig:alphaBeta} plots the performance of NEXT by varying $\beta$ and $\alpha$ values after fixing $\alpha=0.3$ and $\beta=0.2$ respectively.
An obvious observation is that the performance of NEXT starts decrease as either $\alpha$ or $\beta$ increases towards $1.0$. The optimal range of $\beta$ is $[0.1,0.3]$. Also, the optimal range of $\alpha$ is $[0.3,0.6]$. We argue that the meta-data information associated with the users could be more useful on \textit{CA} dataset. Overall, the experimental results demonstrate that the proposed NEXT is competent to exploit the auxiliary meta-data for better recommendation accuracy.

\begin{table}
\centering
\caption{Impact of incorporating auxiliary meta-data in NEXT.}
\label{tbl:metadata}
\begin{tabular}{c||ccc|c}
\toprule
 Meta-data & Acc@1 & Acc@5 & Acc@10 & MAP \\
\midrule
  - & $0.1007$ & $0.2173$ & $0.2793$ & $0.1615$ \\
  \checkmark & \textbf{0.1115} & \textbf{0.2396} & \textbf{0.3038} & \textbf{0.1772}\\
 \bottomrule
\end{tabular}
\end{table}

\eat{
The direct manifestation of Tobler's first law is the sequential relations between POIs.

the first factor affects a user's movement from POI $q_a$ to POI $q_b$ is the geographical distance. She will choose to go to a place near $q_a$.
}

\section{Conclusions}\label{sec:con}
In this paper, we propose a simple neural network framework for next POI recommendation, named NEXT. NEXT derives the spatial intent for a user by calculating POI-based intent and user-based intent separately based on two individual RELU nonlinearities. Under this framework, we can incorporate different contextual factors to enhance next POI recommendation in an unified architecture. Specifically, we incorporate two kinds of temporal context to enhance the intent calculation process. Furthermore, we adopt DeepWalk to encode the spatial constraints such as geographical information and sequential relations pattern into POI embeddings through a pre-training scheme. The experimental results over the three real-world datasets show that the proposed NEXT outperforms  existing state-of-the-art alternatives in terms of MAP and Acc@K. We further show that NEXT achieves better performance in the task of cold-start user recommendation and provide the semantic interpretability for the intent dimenions. This uniqueness makes NEXT an preferrable choice in real-world applications. As a future work, we plan to introduce the attention mechanism into NEXT for better recommendation accuracy.

\section*{Acknowledgment}\label{sec:ack}
This research was supported by National Natural Science Foundation of China (No.~61502344, No.1636219, No.U1636101), Natural Scientific Research Program of Wuhan University (No.~2042017kf0225, No.~2042016kf0190), Academic Team Building Plan for Young Scholars from Wuhan University (No.~Whu2016012) and Singapore Ministry of Education Academic Research Fund Tier 2 (MOE2014-T2-2-066). Chenliang Li is the corresponding author.


\bibliographystyle{apacite}
\bibliography{ref}


\end{document}